\documentclass[conference,10pt]{IEEEtran}
\usepackage[T1]{fontenc}

\usepackage{amssymb,amsmath,amsfonts,amsthm}
\usepackage{bm}
\usepackage{mathrsfs}
\usepackage{cite}
\usepackage[noabbrev,capitalise]{cleveref}
\usepackage{array}
\usepackage{tabularx}
\usepackage[table]{xcolor}
\usepackage{multicol, multirow}
\usepackage{color}
\usepackage{mathtools}
\usepackage{subcaption}
\usepackage{float}
\usepackage{mathtools}
\usepackage{graphicx}
\usepackage{bbm}
\usepackage[utf8]{inputenc}
\usepackage[UKenglish]{babel}
\newcommand{\CLASSINPUTbottomtextmargin}{1in}

\usepackage{algorithm}
\usepackage{algorithmic}
\renewcommand{\IEEEQED}{\IEEEQEDopen}

\newcommand{\reals}{\mathbb{R}}
\newcommand{\complexes}{\mathbb{C}}

\newcommand{\hermitian}{^{\mathsf{H}}} 

\newcommand{\set}[1]{\left\{#1\right\}}

\newcommand{\norm}[1]{\left\lVert #1 \right\rVert}
\newcommand{\abs}[1]{\left|#1\right|}

\newcommand{\CN}[2]{\mathcal{CN}\left( #1,#2 \right)}

\newcommand{\Support}[1]{\ensuremath{\mathop{supp(#1)}}}

\newcommand{\SPARSE}[2]{{\text{SPARSE}_{#1}}\left(#2\right)}

\newcommand{\Userset}{\mathcal{K}}
\newcommand{\Dataset}{\mathcal{D}}
\newcommand{\NumUsers}{K}

\newcommand{\transmittedlength}{L}

\newcommand{\lossfunc}[2]{l\left(#1;#2\right)}
\newcommand{\param}{\bm{\theta}}
\newcommand{\paramopt}{\param^*}
\newcommand{\paramrec}{\hat{\param}}
\newcommand{\Emploss}{F}
\newcommand{\learnrate}{\alpha}

\newcommand{\weight}{w}
\newcommand{\sig}{\bm{s}}
\newcommand{\channelcoef}{h}
\newcommand{\noise}{n}
\newcommand{\noisevec}{\bm{\noise}}

\newcommand{\y}{y}
\newcommand{\yvec}{\bm{\y}}

\newcommand{\x}{x}
\newcommand{\xvec}{\bm{\x}}

\newcommand{\tstep}[1]{^{[#1]}}
\newcommand{\BatchSymb}{\mathcal{B}}
\newcommand{\EpochSymb}{\mathcal{E}}

\newcommand{\measmatrix}{A}
\newcommand{\Lsymb}{\mathcal{L}}
\newcommand{\topl}{top-\ensuremath{\Lsymb{}}}

\newcommand{\Lsparse}{\ensuremath{\Lsymb{}}-sparse}

\newcommand{\sigfull}{\sig^{}}
\newcommand{\sigsparse}{\sig^{\text{sp}}}
\newcommand{\sigiht}{\tilde{\sig}}
\newcommand{\sigrec}{\hat{\sig}}

\newcommand{\Powersymbol}{P}
\newcommand{\Ptot}{\Powersymbol_{\text{tot}}}

\IEEEoverridecommandlockouts
\ifCLASSINFOpdf
\else
\fi
\usepackage[cmintegrals]{newtxmath}

\title{Over-the-Air Federated Learning with Compressed Sensing: Is Sparsification Necessary?}

\author{Adrian Edin and Zheng Chen\\
	Department of Electrical Engineering, Link\"{o}ping University, Sweden\\ Email: \{adrian.edin, zheng.chen\}@liu.se
	\thanks{This work was supported in part by Zenith, Excellence Center at Link\"{o}ping - Lund in Information Technology (ELLIIT), Swedish Research Council (Vetenskapsrådet), and Wallenberg AI, Autonomous Systems and Software Program (WASP) funded by the Knut and Alice Wallenberg Foundation.}
}

\begin{document}
	\maketitle

	\begin{abstract}
		Over-the-Air (OtA) Federated Learning (FL) refers to an FL system where multiple agents apply OtA computation for transmitting model updates to a common edge server. Two important features of OtA computation, namely linear processing and signal-level superposition, motivate the use of linear compression with compressed sensing (CS) methods to reduce the number of data samples transmitted over the channel. The previous works on applying CS methods in OtA FL have primarily assumed that the original model update vectors are sparse, or they have been sparsified before compression. However, it is unclear whether linear compression with CS-based reconstruction is more effective than directly sending the non-zero elements in the sparsified update vectors, under the same total power constraint. In this study, we examine and compare several communication designs with or without sparsification. Our findings demonstrate that sparsification before compression is not necessary. Alternatively, sparsification without linear compression can also achieve better performance than the commonly considered setup that combines both.
	\end{abstract}

	\begin{IEEEkeywords}
		Over-the-Air computation, federated learning, sparsification, compressed sensing, iterative hard thresholding
	\end{IEEEkeywords}

	\section{Introduction}
	Federated Learning (FL) is a distributed machine learning (ML) approach that allows for collaborative training of a common ML model across multiple (possibly massive) agents/devices with local data \cite{mcmahan2017communication}.
	The training process relies on iterative exchange of model updates between local devices and a parameter server (PS).
	The communication bottleneck is a main issue in FL, especially for FL over wireless networks, since the communication resource limitation will greatly affect the learning performance and training latency. For this reason, communication-efficient methods for model update aggregation in FL have attracted wide attention over the past few years \cite{chen2021jointlearnincommunication, async-hu}.  Over-the-Air (OtA) computation has emerged as a promising solution for efficient data aggregation and computation over networks by exploiting the signal superposition property in wireless channels \cite{sahin2023survey, chen2022over}.
	OtA computation relies on simultaneous transmission of data signals from multiple source nodes with appropriate pre-processing and post-processing functions to perform aggregation in the air \cite{goldenbaum2013interferencecalculation}. Many recent works have considered applying OtA computation in FL for aggregating model updates from distributed devices \cite{sery2021cotaf}.

	Even though OtA computation has many advantages in communication efficiency, the increasing number of parameters in current ML models motivates the usage of compression techniques to reduce the amount of data transmitted \cite{alistarh2017qsgd}.
	Since OtA computation relies on linear processing of data before transmission and after reception, compression methods that maintain this linearity is preferred.
	When performing linear compression of a high-dimensional sparse vector, we can use compressed sensing (CS) techniques to reconstruct the original sparse vector with high accuracy from the compressed signal \cite{Eldar2012CS, leinonen2019csmonograph}.
	Combining CS with FL has been investigated in several existing works with different transmission schemes (digital or OtA) for local update aggregation. Typically, a four-step process is adopted: 1) sparsification, 2) compression, 3) transmission, and 4) reconstruction.
	With digital transmission, each device can independently select its sparsification mask, which allows for independent selection of the largest elements  \cite{oh2021communicationefficient,jeon2020compressive, li2021fedcs}. When using OtA computation, ideally the same sparsification mask should be used across different devices to avoid altering the underlying statistics of the aggregated model updates \cite{amiri2020fedfading,optimal-MIMO}.

	In this work, our goal is to investigate the effectiveness of CS-based model compression and reconstruction techniques in OtA FL systems. To this end, we consider several possible communication designs that use sparsification and/or linear compression and compare their performance in terms of learning accuracy, convergence speed, and communication efficiency.
	Our results show that combining sparsification and linear compression might not be an effective strategy. With known sparsity pattern, the usage of CS for model update compression does not bring any obvious advantage as compared to direct transmission of sparsified update vectors with reduced dimension. If one has to apply CS for update compression, then sparsification before compression is not necessary, due to the inherent sparsity structure of aggregated local updates (gradients) in FL.
	\section{System Model}
	We consider an OtA FL system with \(\NumUsers\) devices that collaborate in training an ML model assisted by a central PS for periodic model distribution and aggregation. The set of devices is denoted by \(\Userset = \set{1,\hdots,\NumUsers}\). Each device $k$ holds a local dataset \(\Dataset_k\).
	The total dataset is defined as the combination of all local datasets, i.e.,  \(\Dataset = \bigcup_{k\in\Userset}\Dataset_k\), with \(\Dataset_i \cap \Dataset_j = \emptyset \text{ for any}\,\, i\neq j \).

	For an ML model parameterized by \(\param\in\reals^d\), the goal of training is to find an optimal model parameter vector  \(\paramopt\) that minimizes the global objective function defined as\vspace{-0.1cm}
	\begin{equation}\label{eq:total emploss}
		\Emploss(\param) = \frac{1}{\abs{\Dataset}}\sum_{d\in\Dataset}\lossfunc{d}{\param},
	\end{equation}
	where \(\lossfunc{d}{\param}\) is the per-sample loss function evaluated on sample \(d\).
	Equivalently, we can define a local objective function \(\Emploss_k(\param)\) as the local empirical loss function evaluated on the local dataset \(\Dataset_k\). Then, \eqref{eq:total emploss} can be reformulated as the weighted sum of local objective functions, e.g.,
	\begin{equation}\label{eq:distributed emploss}
		\Emploss(\param) = \sum_{k\in\Userset} \weight_k \Emploss_k(\param),
	\end{equation}
	where the weight \(\weight_k = \frac{\abs{\Dataset_k}}{\abs{\Dataset}}\) indicates the proportion of training data held by device k.

	The most commonly used FL algorithm is the Federated Averaging (FedAvg) \cite{mcmahan2017communication}, which combines stochastic gradient decent (SGD) with local iterations at distributed devices and server-based synchronization of the global model. Each iteration of FedAvg is referred to as one communication round, and in the \(t\)-th round the following steps are executed:
	\begin{enumerate}
		\item The PS transmits the current model \(\param\tstep{t}\) to all devices.
		\item Using the local dataset \(\Dataset_k\), each device runs a certain number of local SGD iterations on some randomly selected mini-batches with a batch size of \(\BatchSymb\).
		\item Each device transmits the local model update \(\Delta\param_k\tstep{t}=\param\tstep{t+1}_{k}-\param\tstep{t}\) to the PS.
		\item The PS computes the weighted average of the received updates to obtain a new global model for the next round
		\begin{equation} \label{eq:model aggregation}
			\param\tstep{t+1} = \param\tstep{t} + \sum_{k\in\Userset} \weight_k\Delta\param_k\tstep{t+1}.
		\end{equation}
	\end{enumerate}

	\subsection{OtA Computation for Efficient Data Aggregation}
	In an FL system, the communication goal in every round is to compute the weighted average of model updates from distributed devices.
	This can be achieved by using OtA computation, a joint communication and computation method originated from the notion of distributed computation of nomographic functions over a multiple access channel (MAC) \cite{nomographic_function}.

	A general nomographic function of $K$ variables can be written as  \vspace{-0.1cm}
	\begin{equation}
		f(\sig_1,\dots,\sig_{\NumUsers})=\varphi\left(\psi_1(\sig_1)+\psi_2(\sig_2)+\ldots+\psi_{\NumUsers}(\sig_{\NumUsers})\right),
	\end{equation}
	where \(\psi_k(\cdot)\) and \(\varphi(\cdot)\) are real-valued continuous functions.

	For our system, let \(\sig_k\in\complexes^{\transmittedlength}\) represent the model update vector from device $k$, we can use \(\psi_k(\cdot)\) as a pre-processing function at the device side before transmission, and  \(\varphi(\cdot)\) as a post-processing function at the PS.\footnote{The original real-valued model update vector can be split into two vectors. Using two orthogonal basis for signal transmission, these two vectors can be viewed as the real and imaginary parts of complex-valued baseband signals.} Then, considering the effect of channel fading and additive noise, the computed function at the PS is
	\vspace{-0.2cm}
	\begin{equation}
		\hat{f}(\sig_1,\dots,\sig_{\NumUsers}) = \varphi\left(\sum_{k\in\Userset}\psi_k(\sig_k)\channelcoef_k+ \noisevec\right).
	\end{equation}
	Here, $h_k\in\complexes$ is the channel gain from device $k$ to the PS, and \(\noisevec\in\complexes^{\transmittedlength}\) is the noise vector where each element follows $\CN{0}{\sigma^2}$.
	Ideally, we want the computed function  $\hat{f}(\sig_1,\dots,\sig_{\NumUsers})$ at the PS to be as close as possible to the following weighted sum
	\begin{equation}
		f(\sig_1,\dots,\sig_{\NumUsers}) = \sum_{k=1}^{\NumUsers}\weight_k\sig_k.
	\end{equation}

	A common choice of the pre-processing function is based on the concept of channel inversion, i.e., we can use
	\begin{equation}
		\psi_k(\sig_k) = \sig_k\cdot \frac{\eta{}\weight_k}{\channelcoef_k},
	\end{equation}
	where \(\eta\) is an amplitude scaling factor, which needs to be adjusted to satisfy some power constraints.
	We assume that the transmission of each element in \(\sig_k\) consumes one channel use, and that the transmission of the entire update vector is under a fixed power limit \(\Ptot\) in every communication round. Then the power constraint gives
	\begin{equation}
		\norm{\psi_k(\sig_k)}^2 \leq {\Ptot}.
	\end{equation}
	For the transmission of each element, this corresponds to a per-symbol power constraint $\frac{\Ptot}{L}$.
	To satisfy the power constraints at all devices, the amplitude scaling factor $\eta$ needs to be
	\begin{equation} \label{eq:ota:eta}
		\eta = \sqrt{\Ptot}\cdot{}\min_{k \in\Userset}{\left\{\frac{|\channelcoef_{k}|}{\weight_{k}\sqrt{{\norm{\sig_k}^2}}}\right\}}.
	\end{equation}

	At the PS side, the post-processing function is simply linear scaling by the factor $1/\eta$. Let $\yvec = \sum_{k \in\Userset}\channelcoef_{k}\psi_k(\sig_{k}) + \noisevec$ be the received signal vector at the PS, then the estimated computation function is
	\begin{equation}\label{eq:received signal nomographic}
		\hat{f}(\sig_1,\dots,\sig_{\NumUsers})=\frac{\yvec}{\eta}= \sum_{k\in\Userset}\sig_{k}\weight_{k} + \frac{\noisevec}{\eta},
	\end{equation}
	which is our desired computation result plus some effective noise with variance  \(\frac{\sigma^2}{\eta^2}\). Since \(\eta\) is often limited by the worst-channel devices, it might be beneficial to set a threshold on the channel gain and drop users with bad channels temporarily. This truncation design can reduce the effective noise variance, but introduce extra bias in the computed function value.

	\subsection{Application of CS in OtA FL}\label{sec:model:cs}
	A vector \(\sig\) is said to be \Lsparse{} if \(\norm{\sig}_0 \leq \Lsymb{}\), i.e. no more than \(\Lsymb{}\) elements are possibly non-zero.
	The support \(\Support{\sig}\) is the set of \(\Lsymb=\abs{\Support{\sig}}\) indices in \(\sig\) where \(\abs{s_i}\geq0, i\in\Support{\sig}\), and \(\abs{s_i}=0 , i\notin\Support{\sig}\).
	Exploiting the sparsity property of a signal can be used for reconstructing a high-dimensional signal from a low-dimensional measurement using CS techniques.
	For a \Lsparse{} vector \(\sigsparse\in\complexes^N\), we can reduce its dimension by a matrix multiplication \(\sigiht = A\sigsparse\), where \(\sigiht\in\complexes^M\) and \(\measmatrix\in\complexes^{M\times N}\) with $M<N$.
	The approximate reconstruction of the original sparse vector \(\sigsparse\) is formulated as
	\begin{equation} \label{eq:iht:minimization_formulation}
		\min_{\sigrec} \norm{\yvec-\measmatrix\sigrec}_2^2,\, \text{s.t.} \norm{\sigrec}_0 \leq \Lsymb{},
	\end{equation}
	where \(\yvec = \sigiht + \noisevec\) and $\noisevec$ is the measurement noise.

	\begin{figure}[t!]
		\begin{algorithm}[H]
			\algsetup{indent=1em}
			\begin{algorithmic}[1]
				\STATE \(\xvec\tstep{0} = \bm{0} \)
				\FOR{ $i = 0,1,\dots$ }
				\STATE $\xvec\tstep{i+1} = H_{\Lsymb{}}\left(\xvec\tstep{i} + \measmatrix\hermitian(\yvec-\measmatrix\xvec\tstep{i})\right)$
				\IF{$\norm{\xvec\tstep{i+1}-\xvec\tstep{i}}^2 < \epsilon$}
				\RETURN $\xvec\tstep{i+1}$
				\ENDIF
				\ENDFOR
			\end{algorithmic}
			\caption{IHT algorithm. $H_{\Lsymb{}}(\bm{a})$ refers to a thresholding operator that sets all but the \(\Lsymb\) largest elements (in magnitude) of $\bm{a}$ to zero.}
			\label{alg:IHT}
		\end{algorithm}
	\end{figure}

	Finding \(\sigrec\) in \eqref{eq:iht:minimization_formulation} can make use of a wide variety of algorithms which have their own benefits, and pose restrictions on the structure of \(\sig\) and \(\measmatrix\) \cite{leinonen2019csmonograph}.
	One common algorithm based on the \(\ell_0\)-norm is iterative hard thresholding (IHT) \cite{blumensath2009IHT}, described in \cref{alg:IHT}. IHT poses restrictions on \(\measmatrix\) to fulfil the Restricted Isometry Property (RIP), which states that for all  \Lsparse{} vectors \(\sigsparse\)
	\begin{equation}
		\left(1-\delta_{\Lsymb}\right)\norm{\sigsparse}^2_2 \leq \norm{\measmatrix\sigsparse}^2_2 \leq \left(1+\delta_{\Lsymb}\right)\norm{\sigsparse}^2_2,
	\end{equation}
	where \(\delta_{\Lsymb}<1\) is the restricted isometry constant.
	Sampling the elements of \(\measmatrix\) from a normal distribution has a high probability of satisfying the RIP.
	Note that \(M\) is a design parameter associated to the compression level \(M/N\). A rule of thumb is to use \(M>\Lsymb\), and preferably \(M\gg\Lsymb{}\), ensuring enough information to accurately reconstruct \(\sigrec\) \cite{blumensath2009IHT}.

	For non-sparse signals, we can obtain its sparse approximation by artificially setting some elements to zero, where the quality of the sparse approximation depend on how the sparsification was performed.
	For transmitting ML model updates, sparse approximations by preserving the \(\Lsymb\) largest elements (known as top-\(\Lsymb\) sparsification) has been numerically shown to have little effect on overall performance when the model updates are aggressively compressed  \cite{alistarh2018convergence}.

	In current literature, applying CS in FL for model compression and reconstruction has been explored under different contexts, depending on whether OtA computation or digital transmission is used for model aggregation \cite{amiri2020fedfading}. These two transmission schemes pose different constraints on the sparsification design. With digital transmission, the model update from each device is processed separately, allowing different sparsification masks to be used at different devices.
	With OtA computation, the received data samples need to be aligned to perform element-wise scaling and superposition. This normally requires that all devices should use the same sparsification mask in order not to change the statistics of the aggregated data.
	In this setup, since each device is only aware of its own update \(\Delta\param\tstep{t}_{k}\), selection of the \(\Lsymb\) largest elements in the aggregated update \(\Delta\param\tstep{t} = \sum_{k\in\Userset}\weight_k\Delta\param\tstep{t}_{k}\) is not possible.
	Instead, if each device use \topl{} sparsification,
	the received compressed signal will contain between \(\Lsymb\) and \(\NumUsers\cdot\Lsymb\) non-zero elements. This method has been numerically tested in \cite{amiri2020fedfading,li2021fedcs} with promising results.
	However, due to the modified statistics in the aggregated model updates which cannot be modeled as additive ``noise'', the effect of this non-identical sparsification design remains to be thoroughly analyzed.
	If we impose all devices to use an identical sparsification mask (i.e. \(\Support{\sigsparse_i} = \Support{\sigsparse_j}, \forall i,j\in\Userset\)), one possible way to construct such a mask is by uniformly random selection of the preserved elements.

	\section{Communication Designs for OtA FL with Sparsity and/or Compression }\label{sec:system}
	In this section, we present four cases of communication design that use different combinations of sparsity and/or compression. Each design introduces different sources of uncertainty and inaccuracy that might cause performance loss in model aggregation, which will be discussed at the end of this section.

	The real-valued model update vector \(\Delta\param_k\in\reals^{d}\) can be transformed into its complex baseband representation \(\sigfull_k\in\complexes^{N}\), with \(N= \lceil \frac{d}{2}\rceil\). The inverse mapping exists at the receiver side to transform the computed function \(\hat{\boldsymbol{f}}_{\text{ag}}\in\complexes^N\) back into real-valued estimated  update vector \(\Delta\paramrec\in\reals^{d}\). \cref{tab:system:notation} contains the description of notations used in this section.

	\begin{table}[htbp]
		\centering
		\renewcommand{\arraystretch}{1.2}
		\begin{tabular}{|p{3cm}|p{4.7cm}|} \hline
			\textbf{Definition} & \textbf{Explanation}  \\ \hline
			\(\sigfull \in\complexes^N\)& Original update vector \\ \hline
			\(\sigsparse =\!\SPARSE{\Lsymb}{\sigfull}\!\in\!\complexes^N\) & \Lsparse{} approximation of \(\sigfull\)  \\ \hline
			\(\boldsymbol{s}^{\text{spr}}\in\complexes^\Lsymb\)& Possibly non-zero elements of \(\sigsparse\)  \\ \hline
			\(\sigiht = \measmatrix\sigsparse\in\complexes^M\) & Compressed version of \(\sigsparse\)\\ \hline
			\(\tilde{\boldsymbol{f}}_{\text{ag}}\in\complexes^M\) & Estimated aggregated compressed update\\ \hline
			\(\hat{\boldsymbol{f}}_{\text{ag}}\in\complexes^N\) & Estimated aggregated update\\ \hline
		\end{tabular}
		\caption{Summary of variables used in the system design.}\label{tab:system:notation}
	\end{table}

	\subsection{Case 1: Direct Transmission of Uncompressed Update}\label{sec:system:raw}
	\begin{figure}[htbp]
		\centering
		\makebox[\columnwidth][c]{
			\includegraphics{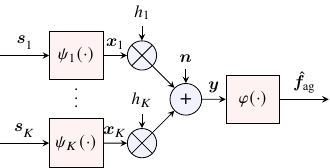}
		}
		\caption{Block diagram of Case 1. Neither compression nor sparsification is used.}
		\label{fig:system:raw}
	\end{figure}

	First, we consider the baseline design where each device simply transmits the full update vector without any compression or sparsification, which we refer to as an uncompressed update. The block diagram for this system is illustrated in \cref{fig:system:raw}. In every round, the transmission of the full update vector consumes \(N\) channel uses, which means that the per-symbol power budget is $\frac{\Ptot}{N}$.

	\subsection{Case 2: Direct Transmission of Uncompressed Sparsified Update}\label{sec:system:only sparse}
	\vspace{-0.2cm}

	\begin{figure}[H]
		\centering
		\makebox[\columnwidth][c]{
			\includegraphics{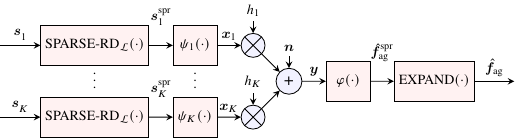}
		}
		\caption{Block diagram of Case 2. Sparsification with no compression. }
		\label{fig:system:only sparse}
	\end{figure}
	In a second design, each device $k$ sparsifies its update vector using the same sparsification mask and keeps only the \({\Lsymb}\) possibly non-zero elements to be transmitted. This operation is marked as \(\text{SPARSE-RD}_{\Lsymb}\) in the block diagram shown in \cref{fig:system:only sparse}. The sparsified vector $\boldsymbol{s}^{\text{spr}}_k$ has reduced dimension and contains only the \(\Lsymb\) possibly non-zero elements. As a result, each transmission round consumes only \(\Lsymb\) channel uses, meaning that the per-symbol power budget is $\frac{\Ptot}{\Lsymb{}}$.

	Note that for this design to work, all devices (including PS) must have knowledge of \(\Support{\hat{\boldsymbol{f}}^{\text{spr}}_{\text{ag}}}=\Support{\sig^{\text{spr}}_k}\) (location information of the preserved elements) to insert the aggregated non-zero elements to the correct positions. At the PS, the operation \(\text{EXPAND}\) maps the aggregated update with reduced dimension \(\hat{\boldsymbol{f}}^{\text{spr}}_{\text{ag}}\in\complexes^\Lsymb\) to a full size update vector \(\hat{\boldsymbol{f}}_{\text{ag}}\in\complexes^N\) by inserting zeros in corresponding positions.

	\subsection{Case 3: Linear Compression with Sparsified Update} \label{sec:system:compression and sparse}

	\begin{figure}[H]
		\centering
		\makebox[0.9\columnwidth][c]{
		\includegraphics{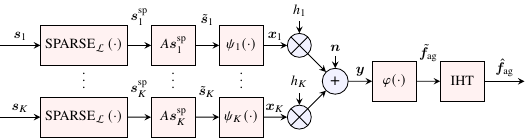}
		}%
		\caption{Block diagram of Case 3. Sparsification and linear compression prior to transmission, with IHT reconstruction at the PS.}
		\label{fig:system:compression and sparse}
	\end{figure}

	This method presents the conventional way of using CS with OtA FL, which performs sparsification prior to compression. The operation for constructing an \Lsparse{} approximation \(\sigsparse_k\) from the full update vector \(\sigfull_k\) is denoted as \(\SPARSE{\Lsymb}{\sigfull_k} = \sigsparse_k\). The sparsification (selection of the \(\Lsymb{}\) elements) is done either by preserving the largest elements at each device or by uniformly random selection.  The linear compression step will reduce the dimension of the sparsified update vector from \(N\) to \(M\) elements in the compressed data vector, which consumes $M$ channel uses for its transmission. The per-symbol power budget is  $\frac{\Ptot}{M}$.

	The PS applies IHT to reconstruct the aggregated sparse model update vector.
	Some side information on how to generate \(\measmatrix\), and possibly the sparsification mask, need to be communicated between the PS and the devices.
	The block diagram of this design is described in \cref{fig:system:compression and sparse}.

	\subsection{Case 4: Linear Compression without Sparsification} \label{sec:system:only compression}
	\begin{figure}[H]
		\centering
		\makebox[\columnwidth][c]{
			\includegraphics{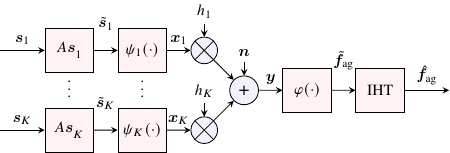}
		}
		\caption{Block diagram of Case 4. Linear compression performed on the original update vector prior to transmission and IHT reconstruction at the PS.}
		\label{fig:system:only compression}
	\end{figure}

	In the last design, we omit the sparsification step and perform linear compression directly on the full update vector, as illustrated in \cref{fig:system:only compression}.
	Same as in Case 3, the compressed data vector contains $M$ elements and the per-symbol power budget is  $\frac{\Ptot}{M}$. The PS applies IHT to reconstruct an $\Lsymb{}$-sparse approximation of the aggregated full update vector $\boldsymbol{f}_{\text{ag}}=\sum_{k\in\Userset}w_k \sigfull_k$, which is not necessarily sparse. The problem can be formulated as
	\begin{equation}
		\min_{\hat{\boldsymbol{f}}_{\text{ag}}} \norm{\yvec-\measmatrix\hat{\boldsymbol{f}}_{\text{ag}}}_2^2,\, \text{s.t.} \norm{\hat{\boldsymbol{f}}_{\text{ag}}}_0 \leq \Lsymb{}.
	\end{equation}
	Note that here the sparsity constraint $\Lsymb{}$ is an artificially chosen parameter that can affect the performance of the reconstruction algorithm.
	The PS only needs to share information about the measurement matrix \(\measmatrix\) to the devices.

	\subsection{Sources of Uncertainty and Inaccuracy}\label{sec:noise sources}
	In the aforementioned  designs, we have several components that can affect the accuracy of the reconstructed aggregated model updates at the PS: the sparse approximation of the update vector, the channel noise, and the reconstruction error in IHT algorithm.
	We need to jointly consider the impact of these different sources of ``noise'' on the aggregation error, and eventually, quantify their effects on the learning performance.
	Another important aspect is the impact of the total power constraint and the difference in per-symbol power budget depending on the sparsification and compression scheme adopted in each design.  For example, with a smaller $M$ (more heavily compressed model), each individual symbol transmission can consume more power, which reduces the OtA computation error caused by channel noise. On the other hand, smaller $M$ means that the original information vector is largely under-sampled, which makes the accurate reconstruction more difficult.

	\section{Simulation Results}
	In our simulations, we create a network with \(\NumUsers^{t} = 100\) users, of which \(K=10\) are randomly selected in every round to participate in the training. The channel gain of each user is randomly generated by \(\channelcoef_k \sim \CN{0}{1}\), with a minimum threshold \(\channelcoef_{\text{th}} = 0.01\). The channel noise uses \(\sigma^2 = 1\), i.e. \(\noise\sim\CN{0}{1}\).

	For the ML task, we consider training a  convolutional neural network (CNN) for digit recognition task, using data from the MNIST dataset \cite{mnist}.
	The CNN model has \(d=21820\) parameter, thus \(N=10920\). Each device holds \(\abs{\Dataset_k}=600\) training data samples and the PS holds a separate validation set with \(10^4\) data samples for validating the performance of the trained model. We consider a non-IID data scenario where each device holds at most two out of the ten classes of digits. During local training, every device uses a learning rate \(\learnrate = 0.01\), batch size \(\BatchSymb = 100\) and number of local epoch \(\EpochSymb =1\).

	Throughout all experiments we use a sparsity level of \(\Lsymb = 500\). When linear compression is involved, the measurement matrix \(\measmatrix\) is generated by first sampling each column of \(\measmatrix'\) uniformly from the unit hyper-sphere. Then forming \(\measmatrix = \frac{\measmatrix'}{1.01\norm{\measmatrix'}_{op}}\), where 1.001 is chosen to ensure that \(\norm{\measmatrix}_{op}<1\).

	\begin{figure}[t!]
		\centering
		\makebox[\columnwidth][c]{
			\includegraphics{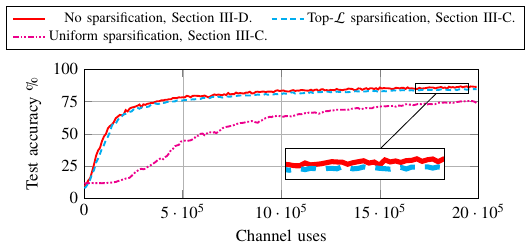}
		}
		\caption{Comparison between the communication designs that use linear compression with no sparsification, sparsification by largest elements and uniform sparsification. The total power budget is $P_{\text{tot}}=10^3$, compressed size \(M=1000\).}
		\label{fig:sparse comp}
	\end{figure}

	\begin{figure}[t!]
		\centering
		\makebox[\columnwidth][c]{
		\includegraphics{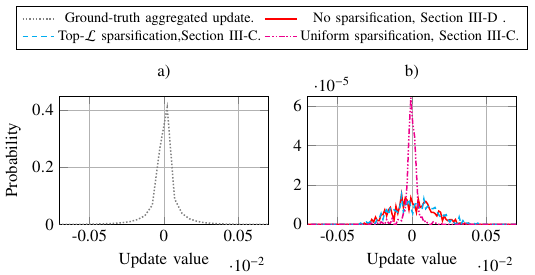}
		}
		\caption{Empirical distribution of aggregate local updates. In b), all distributions have an impulse at $0$ with magnitude \(1-\frac{\Lsymb}{N} (\approx 0.95)\), which is not shown in the plot. The total power budget is $P_{\text{tot}}=10^3$ and the compressed data size is \(M=500\). }
		\label{fig:update distribution}
	\end{figure}
	
	\subsection{Compression with or without Sparsification}

	\cref{fig:sparse comp} shows the performance comparison between three sparsification methods: 1) randomly uniform selection, 2) selection by largest magnitude, and 3) no sparsification, when used in combination with linear compression and reconstruction using IHT. As discussed in \cref{sec:model:cs}, it is unclear how the reconstruction algorithm is jointly affected by the information loss caused by sparsification, reconstruction error caused by the mismatch between the sparsity constraint, and the actual sparsity pattern in the original signal vector.

	In \cref{fig:sparse comp}, we observe that IHT can reconstruct a more accurate update vector when each user applies \topl{} sparsification to its update vector, as compared to uniform sparsification. More importantly, we notice that no sparsification has equal (or better) performance than \topl{} sparsification.

	In \cref{fig:update distribution} we show the empirical distribution of the aggregated local updates, after the trained model reaches 50\% accuracy on the validation set.
	Interestingly, no sparsification method gives very similar distribution as compared to \topl{} sparsification. This suggests that the aggregated update has an inherent (but unknown) sparsity structure that could be used directly for CS-based compression without further sparsification.
	It can also be observed that the result obtained with uniform sparsification appears as a scaled version of the ground-truth distribution, which is expected.
	Another remark is that when using CS-based compression and reconstruction, the average amplitude of model update values is much smaller as compared to the ground-truth aggregated update.

	\subsection{Comparison between Different Communication Designs}
	Here, we compare the performance of the compression without sparsification design (Case 4) in \cref{sec:system:only compression} with the uncompressed designs in \cref{sec:system:raw,sec:system:only sparse}.

	\begin{figure}[t!]
		\centering
		\makebox[\columnwidth][c]{
		\includegraphics{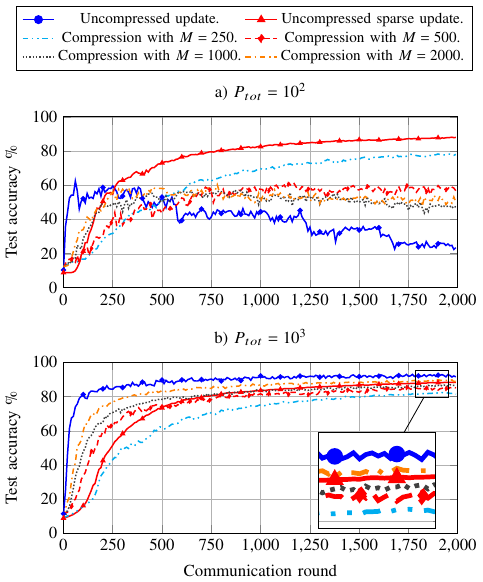}
		}
		\caption{Comparison between transmitting uncompressed updates (both sparse and full update) and compressed updates without sparsification, i.e. from \cref{sec:system:only compression}.}
		\label{fig:M comp round}
	\end{figure}
	
	\subsubsection{Impact of Compression Level}\label{sec:result:impact compression}
	From \cref{fig:M comp round}(b), we observe that the uncompressed update case performs best when the signal-to-noise ratio (SNR) is sufficiently high, e.g., \(\Ptot = 10^3\). With lower SNR, from \cref{fig:M comp round}(a), we see that the uncompressed sparse update case gives a more stable result due to increases per-symbol SNR.

	Comparing the cases with compression but with different values of $M$, we see that with higher SNR in the channel, it is more preferable to use larger $M$ (e.g., $M=2000$ in Fig. 6(b)), while with lower SNR, smaller $M$ (e.g., $M=250$ in Fig. 6(a)) gives better performance. This is mostly caused by the different sources of ``noise'' discussed in \cref{sec:noise sources}. With low SNR in the channel, the channel noise in OtA computation  dominates the inaccuracy of the aggregated model update. With high SNR in the channel, the IHT reconstruction error becomes more important.

	\begin{figure}[ht!]
		\centering
		\makebox[\columnwidth][c]{
		\includegraphics{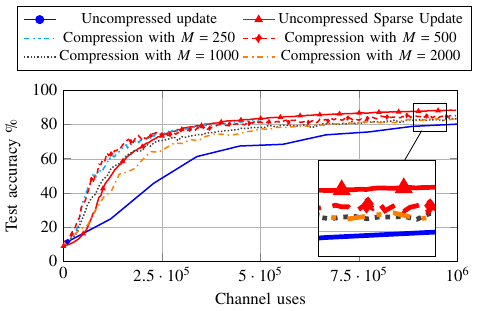}
		}
		\caption{Same as in \cref{fig:M comp round}, but for test accuracy vs. number of channel uses (transmitted symbols). \(P_{tot}=10^3\).}
		\label{fig:M comp normalized}
	\end{figure}
	
	\subsubsection{Impact of Limited Channel Resources} \label{sec:result:impact resources}
	Note that the results in \cref{fig:M comp round} are presented as test accuracy vs. communication round, while each communication round corresponds to different numbers of channel uses for different designs. Here in \cref{fig:M comp normalized}, we compare their performance again by considering test accuracy vs. the number of channel uses. This is particularly important when the communication phase has strict latency requirements. As shown in the figure, the uncompressed update case performs worst in communication efficiency measured by learning performance improvement per channel use. With very high SNR, compression without sparsification achieves the best performance in earlier iterations. In later iterations, the performance becomes comparable to the uncompressed sparse update case.

	\section{Conclusions}
	In this work, we investigated several communication designs for OtA FL systems that use sparsification and/or linear compression techniques along with IHT-based reconstruction of compressed model updates.
	We observed that omitting the sparsification step prior to compression could lead to improved system performance as compared to the common approach that include sparsification.
	Additionally, we explored an alternative scenario where all devices use the same sparsification mask and transmit directly the preserved elements together with their location information to the PS.
	Surprisingly, this sparsification without compression design demonstrated outstanding performance and outperformed the CS-based methods in most cases.

\end{document}